\documentclass[12pt]{article}
\usepackage{amssymb,amsmath,cite,bm}
\usepackage{psfrag,subfigure}
\newtheorem{theorem}{Theorem}
\newtheorem{remark}{Remark}
\newtheorem{proposition}{Proposition}

\title{Non-Minimal Systems with Switching Topology: Dynamics and Controls}

\author{Farhad Aghili}
\date{}
\begin{document}

\maketitle

\begin{abstract}
This paper presents a non-minimal order dynamics model for many analysis, simulation, and control problems of constrained mechanical systems with switching topology by making use of linear projection operator. The distinct features of this model describing dynamics of the dependent coordinates  are: i) The mass matrix $\bar{M}(q)$ is always positive definite even at singular configurations; ii) matrix $\dot{\bar M} - 2 \bar{C}$ is skew symmetric, where all nonlinear terms are lumped into vector $\bar{C}(q, \dot q) \dot q$ after elimination of constraint forces. Eigenvalue analysis shows that the condition number of the constraint mass matrix can be minimized upon adequate selection of a scalar parameter called ``virtual mass'' thereby reducing the sensitivity to round-off errors in numerical computation. It follows by derivation of two oblique projection matrices for computation of constraint forces and actuation forces. It is shown that projection-based model allows feedback control of dependent coordinates which, unlike reduced-order dependent coordinates, uniquely define  spatial  configuration of constrained systems.
\end{abstract}

\section{Introduction}
The system motion of many mechanical systems is often subject to constraints that may arise
from the mechanical interconnections between various bodies of the system, or from the ways in which the system is actuated or its generalized coordinates  are defined \cite{DeLuca-Oriolo-1995}. The unifying modelling approach for simulation and control of such a system is to derive the equations of motion in terms of a minimal set of independent coordinates which  is equal in number to global  degrees-of-freedom exhibited by the constrained system. Although this modelling approach can significantly reduce the number of equations, there are several shortcomings associated with using minimal set of independent coordinates for simulation and control purposes. Firstly, derivation of the minimal-order dynamics model is not always possible. For instance, the rotation motion of a rigid object in three-dimensional space can not be uniquely represented by any three independent variables but at least four dependent coordinates are required. Secondly, mechanical systems might have a time-varying topology, in which forming new independent constraints reduces their number of degrees-of-freedom and therefore a minimal set of independent coordinates with a fixed dimension can not be determined first \cite{GarciadeJalon-Bayo-1994,Blajer-Schiehlen-Schirm-1994,Aghili-2020a}. Consequently, one may need to switch from one minimal order dynamics model to another in a simulation run or to change the structure of a controller whenever the mechanical system changes its number of degrees-of-freedom \cite{McClamorch-Wang-1988,GarciadeJalon-Bayo-1994,Blajer-Schiehlen-Schirm-1994,Aghili-2021a}. Moreover, mechanical systems may occasionally pass through singular configuration during their motion. Singularity occurs when the number of independent constraint equations is reduced in the configuration space, which gives rise to the number of degrees-of-freedom. Consequently, a fixed set of independent coordinates occasionally leads to ill-conditioned matrices when the system changes its topology or the number of degrees-of-freedom \cite{GarciadeJalon-Bayo-1994,Blajer-Schiehlen-Schirm-1994}. Although it is still possible in principle to determine the Lagrangian equations based on minimal coordinates, computation of acceleration becomes extremely sensitive to roundoff errors \cite{Chang-Nikravesh-1985,Lin-Hon-1998,Yoon-Howe-1995}. The augmented Lagrangian formulation proposed in \cite{Bayo-Jalon-1988,Bayo-Ledesma-1996,Cuadrado-Cardenal-Bayo-1997} can handle redundant constraints and singular situations but only through an iterative process. Control of constrained mechanical system based on reduced-order independent coordinates does not allow to control all possible spatial configurations of the system because there is usually no unique relationship between the independent coordinates and the spatial configuration \cite{Aghili-2005,Aghili-Piedboeuf-2003a}.

This paper is the extended version of our earlier work \cite{Aghili-2015c,Aghili-2015b} for non-minimal order dynamics model of mechanical systems based on the notion of linear projection operator. The non-minimal order  model describes dynamics of the dependent coordinates, which unlike independent coordinates, has essentially a fixed dimension even when the mechanical system encounters singularity or changes its number of degrees-of-freedom. The inertia matrix associated with the dependent coordinates $\bar M(q, \mu)$ is expressed in terms of an orthogonal projection matrix, which  defines accessibility of the configuration space, and an arbitrary selected parameter $\mu$ called the virtual mass. The unique aspects of the non-minimal order dynamics model are: i) The constraint inertia matrix $\bar{M} (q, \mu)$ is always positive definite (p.d.) even at  singular configurations; ii) matrix $\dot{\bar M} - 2 \bar C(q, \dot q)$ is skew symmetric. These are analogous to the fundamental properties of dynamics model of unconstrained multibody systems that are widely used in analysis and control. Furthermore,  eigenvalue analysis shows that the condition number of the constraint inertial mass can be minimized upon optimal selection of the virtual mass. Subsequently, two oblique projection matrices are introduced for derivation of the constraint forces and actuation forces. This allows control of the dependent generalized coordinates which  uniquely define the spatial configuration of a constrained mechanical system.

\section{Non-minimum Model of Constrained Mechanical Systems}
\subsection{Generalized model in terms of virtual mass}

Dynamics equations of constrained mechanical systems with generalized coordinates represented by vector $q \in \mathbb{R}^n$ subject to a set of $m$ Pfaffian constraints can be described by
\begin{subequations} \label{eq:dynamics_dae}
\begin{align} \label{eq:Mddq}
M(q) \ddot{q}  +    C(q, \dot{q}) \dot{q} &  =  f_g + f_c + f\\ \label{eq:Adq}
A(q) \dot{q} &=   0
\end{align}
where $M(q) \in \mathbb{R}^{n \times n}$  is the inertia matrix,
$C(q, \dot{q}) \in\mathbb{R}^{n \times n}$ contains Coriolis,
centrifugal terms,  $A(q) \in \mathbb{R}^{m \times n}$ is the constraint matrix, $f_g (q)\in \mathbb{R}^n$ is the generalized conservative forces  owing to the gravitational field, $f_c$ is the generalized constraint forces, and $f \in\mathbb{R}^n$ is the vector of input generalized forces owing to actuation forces $u \in \mathbb{R}^k $ acting on the mechanical system, i.e.,
\begin{equation} \label{eq:f=Bu}
f = B(q) u \quad \mbox{and} \quad v = B^T(q) \dot q
\end{equation}
\end{subequations}
Because of the form of equations in \eqref{eq:f=Bu}, the $u$ and $v$ pair generates the same mechanical work as $f$ and $\dot q$ pair does
\begin{equation} \notag
u^T v = f^T B^T v  = f^T \dot q
\end{equation}

It should be noted that $A(q)$ may or may not be a full-rank matrix depending on the existence of redundant constraints. Therefore, matrix $A(q)$  can be rank-deficient at some particular configurations, i.e., $\mbox{rank}(A(q)) =r$ and $r\leq m$, in which case $r$ is indeed a function of the generalized coordinates $q$. The Pfaffian constraints  in \eqref{eq:Adq} specify that any admissible generalized velocity must lie in the null-space of matrix $A\in\mathbb{R}^{m \times n} $, i.e., $\dot q \in {\cal N}(A)$. On the other hand, the generalized constraint force does not generate work and therefore it should be always orthogonal to the generalized velocity
\begin{equation} \label{eq:fc_null_perp}
\dot{q}^T f_c =0  \; \wedge \; \dot q \in {\cal N}(A)   \quad \Rightarrow \quad  f_c  \in  {\cal N}^{\perp}(  A)
\end{equation}
Let $P \in \mathbb{R}^{n \times n}$ represent the {\em
orthogonal projection} onto the null space of $A$, i.e., ${\cal R}(P) = {\cal N}(A)$, $P^2= P$, and $P^T=P$, where ${\cal R}(P)$ is the range space of $P$. Then, the complimentary projection $Q= I - P$ projects vectors onto the  ${\cal N}^{\perp}(A)$, where $I$ is the identity matrix with adequate dimension. For the sake of notation simplicity, we assume ${\cal N}$ and ${\cal N}^{\perp}$ respectively denote ${\cal N}(A)$ and ${\cal N}^{\perp}(A)$ unless otherwise is specified. Now, the kinematic constraint \eqref{eq:Adq} can be equivalently written as $Q\dot{  q} =   0$ and the time-derivative of which becomes
\begin{equation}\label{eq_(I-P)ddq}
Q \ddot{  q} - \dot{  P} \dot{  q} =0
\end{equation}
\begin{remark}
The time-derivative of an orthogonal projection matrix takes the following form
\begin{equation} \label{eq:OmegaP}
\dot P = \Lambda P + P \Lambda^T
\end{equation}
where  $\Lambda = - A^+ \dot A$.
\end{remark}
Derivation of \eqref{eq:OmegaP} is based on {\em Tikhonov regularization} as described in Appendix A. From the fundamental property of pseudo-inverse $A^+ A A^+ = A^+$, one can infer that matrices $\Lambda$ and $P$ satisfy
\begin{equation} \label{eq:PLambda}
P \Lambda = \Lambda^T P = 0
\end{equation}
The projection-based kinematics model  \eqref{eq_(I-P)ddq} together with \eqref{eq:OmegaP} are at the heart of the following dynamics modeling. Notice that since the projection matrix has a fixed dimension equal to the number of dependent coordinates, it is intrinsically  suitable for  a non-minimal dynamics formulation.

Since $f_c \in {\cal N}^{\perp}$, the projection operator $P$ acts as an annihilator for the
constraint forces, i.e., $P f_c =0$. Therefore, the constraint force can be simply eliminated from
equation \eqref{eq:Mddq} if both sides of the equation are
pre-multiplied by $P$, i.e.
\begin{equation} \label{eq:PMddq}
P M \ddot{q} +   P C P \dot{q} = P (f + f_g),
\end{equation}
Substituting the expression of decomposed acceleration $\ddot q= P \ddot q + \dot P \dot q$ from \eqref{eq_(I-P)ddq} into the above equation  yields
\begin{equation} \label{eq:PMPddq}
PMP \ddot q + \big(PCP + PM \dot P \big)  \dot q= P(f + f_g),
\end{equation}
which indicates that only the null-space component of the generalized forces affects the motion of constrained system. However, the above equation cannot be used to compute the generalized acceleration because $PMP$ is a rank deficient matrix and hence not invertible. To be able to uniquely compute the acceleration vector requires taking  the kinematic equation \eqref{eq_(I-P)ddq}  into account. For combining equations \eqref{eq_(I-P)ddq} and \eqref{eq:PMPddq} into one, we pre-multiply equation \eqref{eq_(I-P)ddq} by $\mu$, where $\mu$ is any positive scalar, and then add both sides of the resultant equation with
\eqref{eq:PMddq}. Thus, we arrive at
\begin{subequations} \label{eq:constraint_system}
\begin{equation} \label{eq_EqMotion2}
\bar{M}(q) \ddot{q} + \bar{C}(q, \dot{q}) \dot{q}   = P(f + f_g),
\end{equation}
where
\begin{align} \label{eq:Mbar}
\bar {M} (q) & =   P    M(q)   P + \mu Q,\\ \label{eq:Cbar2}
\bar{C}(q, \dot{q}) & =   P C(q, \dot{q}) P + P M (\Lambda P + P \Lambda^T) - \mu \Lambda P
\end{align}
\end{subequations}
Here \eqref{eq:Cbar2} is obtained by substituting $\dot P$ from \eqref{eq:OmegaP} and using identity $\mu P \Lambda^T \dot q = \mu P \Lambda^T P \dot q =0$, which is inferred from \eqref{eq:PLambda}. Equation \eqref{eq_EqMotion2} constitutes the {\em non-minimal order} dynamics model of mechanical systems in the standard form where $\bar M(q)$ can be treated as the constraint inertia matrix.
\begin{theorem} \label{th:two-preoperties}
\begin{enumerate}
\item Constraint inertia matrix $\bar{M}$ is always symmetric positive-definite (p.d.) even if $A(q)$ becomes singular.
\item Matrix $\frac{d}{dt}{\bar{M}} - 2 \bar{C}$ is  skew-symmetric.
\end{enumerate}
\end{theorem}
{\sc Proof:} For any  non-zero vector $ x \in \mathbb{R}^n$, we can say
\begin{equation}  \label{eq:twoMass}
x^T \bar{M} x  =  x^T_{\parallel} M x_{\parallel} + \mu \| x_{\perp} \|^2 >0,
\end{equation}
where $x_{\parallel}=P x$ and $x_{\perp}=Q x$ are the orthogonal decomposition components, i.e.,  $x=x_{\parallel} \oplus x_{\perp}$. Since $M$ is a p.d. matrix, both terms in RHS of \eqref{eq:twoMass} are positive and hence they can not cancel each other out. Therefore, the summation remains positive definite  unless $x_{\parallel}$ and $x_{\perp}$ are both zero that is not possible because $x\neq 0$. Therefore, $\bar M$ must be a p.d. matrix.

Moreover, using the expression of the matrices involved from \eqref{eq:Mbar} and \eqref{eq:Cbar2} gives
\begin{align} \notag
\dot{\bar{M}} - 2 \bar{C}  = & \dot P M P +  P \dot{M} P +  P M \dot P  - \mu \dot P \\ \notag
&- 2 P C P - 2 P  M \Lambda P - 2PMP \Lambda^T  + 2 \mu \Lambda P \\ \notag
= & P\dot MP - 2PCP + P \Lambda^T MP + \Lambda P M P \\ \notag
& - P M \Lambda P - PMP\Lambda^T + \mu \Lambda P- \mu P\Lambda^T \\ \notag
= & [P(\dot M - 2C)P] + [(P\Lambda^T M P) - (P\Lambda^T M P)^T] \\ \label{eq M-2C}
&+[\Lambda P M P - (\Lambda P M P)^T ] + \mu[\Lambda P  - (\Lambda P)^T]
\end{align}
Knowing the fact that matrices  $P$ and $\dot M - 2C$ are respectively symmetric and skew-symmetric, one can readily verify that all four brackets in the RHS of \eqref{eq M-2C} contain skew-symmetric matrices and therefore  their summation $\dot{\bar{M}} - 2 \bar{C}$ must be a skew-symmetric matrix too. $\Box$

\subsection{Eigenvalue analysis of constraint mass matrix}

The projection matrix is dimensionless and hence in view of \eqref{eq:Mbar}, we can say $\mu$ must have the dimension of mass. From \ref{eq:Mbar}, we get
\begin{equation} \notag
T = \frac{1}{2}\dot{q}^T \bar{M}(q, \mu) \dot{q} = \frac{1}{2}\dot{q}^T {M}(q) \dot{q} \qquad \qquad \forall \mu \in \mathbb{R}_{+}
\end{equation}
Therefore, we call $\mu$ as the {\em virtual mass} of the constrained systems because it has no effect on the kinetic energy. In other words, the virtual mass can take any positive value without affecting the solution of the equations of motion \eqref{eq_EqMotion2}. Yet certain values of $\mu$ are preferable in the sense of achieving minimum {\em condition number} of the constraint inertia matrix. Thus we seek $\mu$ such that
\begin{equation} \label{eq:condM}
\min_{\mu} \mbox{cond}(\bar M(q,\mu))
\end{equation}
\begin{proposition}
The optimal value of $\mu$ minimizing the condition number of the constraint inertia matrix  satisfies the following inequalities
\begin{equation} \label{eq:optimum_mu}
\lambda_{\stackrel{\rm min}{\neq0}}(PMP) \leq \mu \leq \lambda_{\rm max}(PMP)
\end{equation}
\end{proposition}
{\sc Proof:} Consider the characteristic equation of the constraint mass matrix
\begin{equation} \notag
\big( PMP + \mu Q \big) x - \lambda x =0
\end{equation}
Clearly $\lambda=\mu$ is the eigenvalue for all orthogonal eigenvectors which span ${\cal N}^{\perp}$ because $\lambda=\mu$ means $(PMP - P)x =0 \quad \forall x \in {\cal N}^{\perp}$. The remaining set of  orthogonal eigenvectors  must lie in ${\cal N}$ that are  corresponding to the non-zero eigenvalues of $PMP$
\begin{equation} \notag
PMP x - \lambda x =0 \qquad  \lambda \neq 0 \quad \forall x \in {\cal N}
\end{equation}
Therefore, the set of all eigenvalues of the p.d. matrix $\bar M$ is  the union of the above sets corresponding to the eigenvectors in ${\cal N}$ and ${\cal N}^{\perp}$, i.e.,
\begin{equation} \label{eq:lamb(M_bar)}
\lambda(\bar M)=: \big\{\underbrace{\mu, \cdots, \mu}_{r}, \;  \underbrace{ \lambda_{\stackrel{\rm min}{\neq0}} (PMP), \cdots, \lambda_{\rm max}(PMP)}_{n-r} \big\}
\end{equation}
where $\{ \lambda_{\stackrel{\rm min}{\neq0}}(PMP), \cdots, \lambda_{\rm max}(PMP) \}$ are all non-zero eigenvalues of $PMP$. According to \eqref{eq:lamb(M_bar)} the condition number of  $\bar M$, which is simply the ratio of the largest to smallest eigenvalues, is
\begin{equation} \label{eq:cond_M}
\mbox{cond}(\bar M) = \frac{\max(\mu, \lambda_{\rm max}(PMP) )}{\min(\mu, \lambda_{\stackrel{\rm min}{\neq0} }(PMP))}
\end{equation}
Clearly, the RHS of \eqref{eq:cond_M} is at its minimum if $\mu$ is selected according to \eqref{eq:optimum_mu}. $\Box$

\section{Control}
\subsection{Admissible generalized input force and oblique projection}
As far as control of the constrained systems is concern, only the null-space component of the generalized force $f_{\parallel}$ is relevant because the orthogonal complement component of the force $f_{\perp}$ does not contribute to the motion of the system. However, \eqref{eq:f=Bu} stipulates that any admissible vector of generalized forces must satisfy
\begin{equation} \label{eq:R(B)}
f \in {\cal R}(B),
\end{equation}
Hence the generalized forces may inevitably contain components of $f_{\perp}$ in addition to $f_{\parallel}$ to make \eqref{eq:R(B)} happen. In this section, we show that $f$ is indeed related to $f_{\parallel}$ by an oblique projection matrix. To this end,  pre-multiplying both sides of \eqref{eq:f=Bu} by $P$ gives
\begin{equation} \label{eq:PBu}
f_{\parallel} = P B u
\end{equation}
The above equation will have at least one solution for $u$ if ${\cal R}(PB) \subseteq {\cal R}(P) = {\cal N}$. Therefore, one can conclude that the condition for existence of solution for the actuator force vector is that
\begin{equation} \label{eq:Range-PB}
{\cal R}(PB) = {\cal N}, \quad  \mbox{or} \quad {\cal N} \subseteq {\cal R}(B)
\end{equation}
If the above condition is satisfied, then the reciprocal of the relationship \eqref{eq:PBu} is given by
\begin{equation} \notag
u = \Gamma f_{\parallel}  \quad \leftarrow \quad \min \| u \|
\end{equation}
where $\Gamma=(PB)^+ = (B^T P B)^{-1} B^T P$. Consequently the entire vector of the generalized forces  can be computed from
\begin{equation} \label{eq:p1_inverse}
f =  R f_{\parallel},
\end{equation}
where $R$ is an {\em oblique projection matrix}, i.e., $R^2 =R$ and  $R^T \neq R$, given by
\begin{equation} \label{eq:Q_definition}
R:= B \Gamma =B(B^T P B)^{-1} B^T P.
\end{equation}
By virtue of \eqref{eq:Range-PB}, one can conclude that $R$ projects a vector to the range-space of $B$ along the null-space of $A$.  We will explore some useful properties of the oblique projection in the following remark.
\begin{remark}
If \eqref{eq:Range-PB} holds, then the oblique projection $R$ satisfies:
\begin{equation} \label{eq:QP=Q}
PR = P \quad \mbox{and} \quad RP = R
\end{equation}
\end{remark}
See Appendix B for proof. Identities \eqref{eq:QP=Q}  mean that the projection matrix $R$ maps any vector belonging to ${\cal N}$ onto itself. This is an important property  that will be later used for the control design purposes.

\begin{remark} \label{rem:Gamma}
If the form of the $B$ matrix is such that $PB = B$, then $\Gamma = B^+$ and $R= BB^+$ becomes an orthogonal projection matrix.
\end{remark}
\subsection{Control of dependent coordinates}
Suppose the control objective is to regulate the spatial configuration of system \eqref{eq:dynamics_dae} according to the desired $q^*$. Consider the following control law
\begin{equation} \label{eq:control-law}
f = -R(q) \Big( f_g(q) +  K_p   (e+ \sigma \| e \| \eta) +   K_d \dot q \Big)
\end{equation}
where $e = q - q^*$ is the control error,  $K_p >0$ and $K_d>0$ are the positive gains, $\sigma >1$ is a constant scalar, $R(q)$ is the oblique projection matrix as defined in \eqref{eq:Q_definition}, and $\eta$ is a unit vector in the direction of the generalized velocity defined as follow
\begin{equation}
\eta= \left \{ \begin{array}{ll} \frac{\dot q}{\| \dot q \|}   \quad & \mbox{if} \quad \|\dot q \| \neq 0 \\ \xi & \mbox{otherwise} \end{array} \right.
\end{equation}
Here, $\xi$ is  a constant unit vector in ${\cal N}(A)$ -- hence $P \eta = \eta$ and $\| \eta \| =1$. Notice that  control law \eqref{eq:control-law} can be equivalently rewritten so as to produce the actuation forces $u$  if $R(q)$ is replaced by $\Gamma(q)$ in the control law. The closed-loop dynamics under control law \eqref{eq:control-law} becomes
\begin{equation} \label{eq:error}
\bar M \ddot q  = -\bar C \dot q - P R \Big( K_p (e+ \sigma \| e \| \eta) + K_d \dot q \Big).
\end{equation}
Now consider Lyapunov function
\begin{equation} \label{eq:V}
V = \frac{1}{2} \dot{q}^T \bar{M} \dot{q} + \frac{1}{2}   e^T  K_p   e
\end{equation}
Then using the property that matrix $\dot{\bar M} - 2 \bar C$ being skew-symmetric  and identity $PR =P$  in the expression of the time-derivative of $V$ along \eqref{eq:error}, we arrive at
\begin{align*}
\dot{V} & = \dot{q}^T(\frac{1}{2} \dot{\bar{M}} - \bar C) \dot{q} - \dot{q}^T   P K_d \dot q - \sigma \| e \| \dot q^T K_p \eta
 + e^T K_p (I - P) \dot{q} \\
& \leq - \dot{q}^T   K_d \dot{q} - \sigma K_p \| e \| \; \| \dot q \|  <0
\end{align*}
Clearly, we have $\dot V=0$ only if $\dot q =0$. Thus substituting $\ddot{q}=\dot{q}= 0$ in~\eqref{eq:error}, we can find the largest invariant set with respect to system ~\eqref{eq:error} as the following
\begin{equation} \label{eq:Pe}
\dot q =0 \quad \wedge \quad P e + \sigma {\| e \|} \eta =0.
\end{equation}
Therefore, according to LaSalle's Global Invariant Set Theorem \cite{Lasalle-1960},  the solution of system~\eqref{eq:error} asymptotically converges to the invariant set. The magnitude of the first term in the above satisfies $\| P e \| \leq \| e \|$ whereas the magnitude of the second term is $\sigma \|e \|$. Therefore, if we select $\sigma>1$, then the only solution of \eqref{eq:Pe} is $e=0$. The above development is summarized in following theorem
\begin{theorem}
Suppose control law \eqref{eq:control-law} is applied to a constrained system \eqref{eq:dynamics_dae}, where  $K_d>0$, $K_p >0$, $\sigma >1$, and condition \eqref{eq:Range-PB} holds. Then, $q\rightarrow q^*$ and $\dot q \rightarrow 0$ as $t\rightarrow \infty$.
\end{theorem}

\section{Generalized acceleration and generalized constraint forces}
\subsection{Pseudo-inverse of the constraint inertia matrix}
Since $\bar M$ is invertible, the acceleration of the dependent generalized coordinates can be always computed from  \eqref{eq_EqMotion2} regardless the ill-conditioning of the constraint.  By inspection, one can verify
\begin{equation} \notag
\bar{C}(q , \dot q) \dot q = \bar{C}'(q , \dot q) \dot q
\end{equation}
where $\bar{C}'(q , \dot q) = P C(q , \dot q) + (PM - \mu I) \Omega$ and matrix
\begin{equation} \notag
\Omega= \Lambda - \Lambda^T
\end{equation}
is skew-symmetric. It is worth noting that $\Omega$  can be treated  as representation of  angular rates  in the ``$n$-dimensional space''. Moreover, from definition of the constraint inertia matrix we have $\bar M Q =\mu Q$ or equivalently $\mu \bar{M}^{-1} Q =Q$ meaning that matrix $ \bar{M}^{-1} Q $ must be symmetric because $Q=Q^T$. It follows from $\bar{M}^{-1} Q =Q\bar{M}^{-1}$ that
\begin{equation} \label{eq:barM_invQ}
\bar M^{-1} P = P \bar M^{-1}
\end{equation}
Using the fact that matrices $P$ and $\bar M^{-1} $ commute  and identity $Q \Omega = \Omega$, one can compute the vector of generalized acceleration  from
\begin{align} \notag
\ddot q & = \bar{M}^{-1}P\Big(f+ f_g+ C(q, \dot q) \dot q  + M \Omega \dot q \Big) - \mu \bar{M}^{-1} \Omega \dot q  \\ \label{eq:ddq_sim}
& =\bar M^{-1}P \Big(f + h(q, \dot q) \Big) - S^T \Omega \dot q
\end{align}
where vector $h(q, \dot q) = f_g(q) - C(q, \dot q)\dot q$ contains all nonlinear terms associated with the  gravitational, Coriolis, centrifugal forces of the unconstrained system,  and non-symmetric matrix
\begin{equation}  \label{eq:S_definition}
S :=I-M \bar{M}^{-1}P,
\end{equation}
is indeed another oblique projection matrix as shown in the following. By virtue of \eqref{eq:barM_invQ}, we have $\bar M P \bar M^{-1} = \bar M \bar M^{-1} P =P$ leading to $PMP\bar M^{-1}P=P$, which in turn, is used in the following derivation
\begin{equation} \notag
S^2 = I -2 M \bar M^{-1} P + M \bar M^{-1} P M P \bar M^{-1} P = S
\end{equation}
It is worth noting that the matrix product $P \bar M^{-1}$ which appears in \eqref{eq:ddq_sim} and \eqref{eq:S_definition} can be alternatively obtained  from  pseudo-inversion of the constraint inertia matrix when $\mu$ is set to zero, i.e.,
\begin{equation} \notag
\bar M^{-1} P = \bar M_o^+ \quad \mbox{where} \quad \bar M_o = PMP
\end{equation}
See Appendix C for the proof. Thus, $S=I - M \bar{M}^+_o$.

\subsection{Constraint force and oblique projection}
Upon substitution of the acceleration from  \eqref{eq:ddq_sim} into \eqref{eq:Mddq} and rearranging the latter equation, we arrive at the equation the constraint forces in the following compact form
\begin{equation} \label{eq:fc_R}
f_c = -S \big(f + h(q, \dot q) - M \Omega \dot q \big)
\end{equation}
It can be verified that $S$ satisfies
\begin{equation} \label{(I-P)R}
Q S  = S  \quad \mbox{and} \quad   S Q  = Q
\end{equation}
The first identity in the above indicates that ${\cal R}(S)\equiv {\cal N}^{\perp}$, which is the direct consequence of  \eqref{eq:fc_null_perp}. The second equality means that $S$ maps any vector in ${\cal N}^{\perp}$ to itself, and hence by making use of identity  $PSMQ \equiv 0$ we can rewrite \eqref{eq:fc_R} in the form
\begin{equation} \label{eq:fc_fp}
f_{\perp}+ f_c =   - S \big(f_{\parallel} + h(q, \dot q) \big) + Q M \Omega \dot q
\end{equation}
Equation \eqref{eq:fc_fp} is useful for force control application, because \eqref{eq:fc_fp} algebraically determines the ${\cal N}^{\perp}$ component of the generalized input forces to generate desired constraint forces.

\section{Conclusions}
The non-minimal order modeling formulation presented in this paper
opens up the way for  many analysis, simulation, and control
problems of constrained mechanical systems in the presence of
singular configuration. The orthogonal projection matrix $P$
representing the accessible space of the spatial configuration is
the cornerstone of this modeling formulation in the standard form
in pair with unconstraint multibody dynamics model. Although
dependent generalized coordinates is with a dimension higher than
the number of degrees-of-freedom belonging to the constrained
system, the non-minimal order modeling formulation has distinct
features useful for simulation and control purposes particularly
whenever the mechanical system encounters singularity or changes
its number of degrees-of-freedoms: i) The inertia mass matrix
$\bar{M}$ has fixed dimension equal in number of dependent
coordinates and remains always positive definite even at singular
configurations; ii) matrix $\dot{\bar{M}} - 2 \bar{C}(q, \dot q)$
is skew-symmetric. It has been also shown that $\bar M$ could be
parameterized as a function of any arbitrary selected virtual mass
$\mu$ and subsequently the ones minimizing the condition number of
the mass matrix were found through eigenvalue analysis. Two
oblique projection matrices $S$ and $R$ have been introduced. The
former was used to compute directly the generalized constraint
forces without a recourse to derivation of the Lagrange
multipliers, which are  undeterminate  at singularities. The
latter was used to compute  the actuation force for direct control
of the dependent generalized coordinates which, unlike any
independent variables, uniquely define the spatial configuration
of constrained mechanical systems.


\section*{Appendix A}
The projection operator in terms of the pseudo-inverse of $A$ is given by $P= I - A^+ A$ and therefore
\begin{equation} \label{eq:dotPA+}
\dot P= - \frac{d}{dt}A^+ A - A^+ \dot A.
\end{equation}
According to the {\em Tikhonov regularization}, the pseudo-inverse is limit
\begin{equation} \label{eq:Tikhonov}
A^+ = A^T  L\quad  \mbox{where} \quad L= \lim_{\epsilon \rightarrow 0}(A A^T + \epsilon I)^{-1}
\end{equation}
Notice that the limit exists even $(A A^T)^{-1}$ does not exist \cite{Golub-VanLoan-1996}. Using \eqref{eq:Tikhonov} in \eqref{eq:dotPA+} yields
\begin{equation} \label{eq:dotP-A+}
\begin{split}
\dot P & = \lim_{\epsilon \rightarrow 0} -\dot A^T L A + A^T L \dot A A^T L A + A^T L A \dot A^T L A - A^+ \dot A \\
&= A^+ \dot A A^T A^{+T} + A^+ A \dot A^T A^{+T} - \dot A^T A^{+T} - A^+ \dot A \\
&= ( A^+ \dot A)(A^T A^{+T} -I) + (A^+ A - I ) \dot A^T A ^{+T} \\
&=  \Lambda P + P \Lambda^T
\end{split}
\end{equation}

\section*{Appendix B}
By definition, we have
\begin{equation} \notag
PR =  PB (PB)^+
\end{equation}
Clearly matrix $PR$ itself must be an orthogonal projection which maps a vector onto the orthogonal complement of the null-space of $(PB)^T$. Therefore, according to  the fundamental theory of linear algebra relating the null space and the range space of a linear operator, we have
\begin{equation} \label{eq:R(PQ)}
{\cal R} (PR) \equiv {\cal N}^{\perp}((PB)^T) \equiv  {\cal R} ((PB))
\end{equation}
It can be inferred from \eqref{eq:Range-PB} and \eqref{eq:R(PQ)} that
\begin{equation}
{\cal R} (PR) \equiv {\cal N}
\end{equation}
meaning that projection matrices $PR$ and $P$ are indeed identical.

\section*{Appendix C}
Since $\mbox{rank}(\bar M) = \mbox{rank}(P) = r$ and matrix $\bar M$ is symmetric, the Singular Value Decomposition \cite{Golub-VanLoan-1996} of $\bar M$ takes the form
\begin{equation}
\bar M = \begin{bmatrix} V_1 & V_2 \end{bmatrix} \begin{bmatrix} \mu I_r & 0 \\ 0 & \Sigma \end{bmatrix} \begin{bmatrix} V_1^T \\ V_2^T \end{bmatrix}
\end{equation}
where $\Sigma=\mbox{diag}\{\lambda_{\stackrel{\rm min}{\neq 0}}(PMP),\cdots,\lambda_{\rm max}(PMP)\}$ are the singular
values according to \eqref{eq:lamb(M_bar)}, $V=[V_1 \;\; V_2]$ is a unitary matrix so that $\mbox{span}(V_1) \equiv {\cal N}^{\perp}$ and $\mbox{span}(V_2) \equiv {\cal N}$. That is
 \begin{equation}
P = V_2 V_2^T, \quad  V_2^T V_2 =I, \quad \mbox{and} \quad V_1^T V_2 =0
\end{equation}
and therefore
\begin{align} \notag
\bar M^{-1} P &= \begin{bmatrix} V_1 & V_2 \end{bmatrix} \begin{bmatrix} \mu^{-1} I & 0 \\ 0 & \Sigma^{-1} \end{bmatrix} \begin{bmatrix} V_1^T \\ V_2^T  \end{bmatrix} V_2 V_2^T \\ \notag &= V_2 \Sigma^{-1} V_2 = M_o^+
\end{align}

\bibliographystyle{IEEEtran}

\end{document}